# WeSSQoS: A Configurable SOA System for Quality-aware Web Service Selection

Oscar Cabrera, Marc Oriol, Xavier Franch, Lidia López, Jordi Marco
Universitat Politècnica de Catalunya (UPC)
Barcelona, Spain
{ocabrera, franch}@essi.upc.edu
{moriol, llopez, jmarco}@lsi.upc.edu

Olivia Fragoso, René Santaolaya
Centro Nacional de Investigación y Desarrollo Tecnológico (CENIDET)
Cuernavaca, México
{ofragoso, rene}@cenidet.edu.mx

*Abstract*— Web Services (WS) have become one the most used technologies nowadays in software systems. Among the challenges when integrating WS in a given system, requirements-driven selection occupies a prominent place. A comprehensive selection process needs to check compliance of Non-Functional Requirements (NFR), which can be assessed by analysing WS Quality of Service (QoS). In this paper, we describe the WeSSQoS system that aims at ranking available WS based on the comparison of their QoS and the stated NFRs. WeSSQoS is designed as an open service-oriented architecture that hosts a configurable portfolio of normalization and ranking algorithms that can be selected by the engineer when starting a selection process. WS' QoS can be obtained either from a static, WSDL-like description, or computed dynamically through monitoring techniques. WeSSQoS is designed to work over multiple WS repositories and QoS sources. The impact of having a portfolio of different normalization and ranking algorithms is illustrated with an example.

## I. INTRODUCTION

Web Services (WS) [1] are a technology that integrates a set of protocols and standards for interchanging data among software applications that are developed using different programming environments and languages, and executed in different platforms. Several existing open standards (remarkably, XML, SOAP, HTTP, WSDL, UDDI…) are ensuring this interoperability.

WS have become a technology of reference in the implementation of any kind of software. This success has triggered the emergence of a huge WS marketplace. Consequently, for a given functionality we may find hundreds of WS that can be selected in several ways. This proliferation of WS increments the chances to find existing software for satisfying the stated needs, but at the same time raises new problems and challenges. Among them we address the problem of selecting the most appropriate WS for a given context of usage [2]. Usually this problem is studied in relation with the requirements elicited from the system stakeholders, in other words, the goal is selecting the WS that satisfies "better" the stakeholder requirements.

Concerning requirements, we consider here the classical distinction among functional and non-functional requirements [3]. With respect to functional requirements, it is necessary to validate that a WS fulfils the functionality expected by stakeholders. On the other hand, non-functional requirements (NFR) refer to the quality of service (QoS) that offers the WS, i.e. behavioural characteristics that the WS exhibits for offering a given functionality: cost, response time, availability, etc. Usually, the expression of NFRs in terms of QoS yields to a certain service level agreement (SLA). Therefore, we can assess how well a WS *w* satisfies an NFR *r* by checking that *w*'s QoS satisfies those clauses from the SLA that refer to concepts inherent to *r*.

Given this context, our work proposes a software system for ranking a given set of WS that belong to a certain software domain. We assume that the functional requirements are used to determine this software domain, therefore our system focuses in the ranking based on the satisfaction of NFRs. Basically, the issues to determine are: how NFRs are expressed; which is the measure of the satisfaction of an NFR in a given WS; how are these individual measures combined in order to rank the WS according to a set of NFRs; how the QoS of a WS may be obtained; where the WS are obtained from.

More specifically, this paper presents WeSSQoS (Web Service Selection based on Quality of Service), a software system for selecting WS based on their QoS and NFRs. WeSSQoS proposes an open service-oriented architecture (SOA) that is able to manage several algorithms for computing the adequacy of a WS with respect to NFRs. NFRs are expressed by means of formulae stated over QoS attributes (i.e., SLA clauses) coming from the quality model proposed in an earlier work [4]. NFRs are classified as mandatory and optional, and this information may be used for ranking the results. WeSSQoS is designed to work over several WS repositories that eventually can be built using different technologies. For knowing the behaviour of the accessible WS with respect to the selection criteria, it is possible to use either the description of the QoS if it is part of the WS definition, or the results of monitoring the WS, obtaining then the real, updated QoS of the WS. In this sense, we share the vision of [5] that proposes to define a priori only static attributes (like cost) whilst dynamic attributes (like response time or availability) should be obtained through monitoring.

The rest of the paper is organized as follows. Section II classifies some existing frameworks according to some criteria. Sections III to V describe the WeSSQoS framework: the selection process, its architecture, its algorithms and the services offered. Sections VI and VII describe a prototype and provide some validation. Finally, section VIII presents the conclusion and future work.

TABLE I. COMPARATIVE TABLE OF THE FRAMEWORKS.

| Proposal | Architectural Style | Attributes | QoS Data Source | Multialgorithm | Multirepository | Prototype available |
|---|---|---|---|---|---|---|
| Al-Masri et al.[5] | CBA | 6 dynamic 3 static | Static attributes defined and dynamic attributes monitored | No | Yes | Yes |
| QCWS [6][7] | CBA | 4 dyn. + 1 sta. | Defined | Yes, closed | No | No |
| Liu et al. [8] | CBA | 1 dyn. + 5 sta. | Static defined and dynamic monitored | No | No | No |
| QoS-IC [9] | CBA+SOA | 5 dyn. + 1 sta. | Defined | No | No | No |
| Wang et al. [10] | Only describes the algorithm | Configurable 1 dyn. + 5 sta. | Defined | No | No | No |
| Maximilien et al. [11] | SOA | Config. 3 dyn. | Monitored | No | No | No |
| Hu et al. [12] | SOA | 3 dyn. + 2 sta. | Defined | No | No | No |
| D' Mello et al. [13] | CBA+SOA | Config. 3 dyn. + 2 sta. | Static defined and dynamic monitored | No | No | No |
| WeSSQoS | CBA+SOA | Config. 9 dyn. + 1 sta. | Static defined and dynamic monitored | Yes | Yes | Yes |

## II. RELATED WORK

There exist several frameworks on ranking and selection of WS accordingly to their QoS. In Table 1 we show a representative sample of these proposals including our WeSSQoS comparing them accordingly to the criteria below:

- *Architectural style*. Architecture in which the framework has been developed. We find Component Based Architectures (CBA), Service Oriented Architectures (SOA) and a combination of both. Adoption of SOA supports the integration in other systems.
- *Attributes*. Quality attributes considered in those systems. In some cases, a small predefined set of quality attributes is being used, whereas other frameworks allow the usage of arbitrary ones (although they may validate the proposal with a given set).
- *QoS data source*. Describes if quality attributes are declared in services definition or for dynamic quality attributes, if their value is obtain through monitoring.
- *Multialgorithm*: Describes if the system is able to work with more than one selection algorithm. Up to our knowledge, only QCWS offers this capability. However, since it is not an SOA, it does not allow adding external algorithms.
- *Multirepository*: Describes if the system is able to obtain the data from different repositories and combine the information to extend the number of services and quality attributes to evaluate. The only system that presents this characteristic is from Al-Masri et al., whose framework obtains the list of web services from several sources (UDDIs, ebXMLs, search engines and service Portals). However, it does not specify a method to combine services data (i.e.: if sources have the same service with different QoS data: cost, brand reputation…).
- *Prototype available*: Although in most of proposals a prototype is being described, and even some of them have a web page (e.g., QCWS), we have only found a tool available from Al-Masri et al. proposal.

## III. WESSQOS: AN OVERVIEW

Fig. 1 shows the WeSSQoS selection process with input:

- WSlist, a QoS matrix of size $k \times n$, where $(w_1, …, w_k)$ are the candidate WS and $(q_1, …, q_n)$ are the quality attributes referred in the NFRs. WSlist$[i, j]$ stands for the value of the quality attribute $q_j$ in the WS $w_i$.
- lreqs, a NFR vector of size $n$, where lreqs$[i]$ specifies (1) the value that is required for the attribute $q_i$, (2) a Boolean value that indicates if the attribute's value is to be minimized or maximized, and (3) another Boolean value that indicates if the required attribute's value is mandatory or not. A value is mandatory when it cannot be higher than required in the case that it has to be minimized, or cannot be lower than required in the case that it has to be maximized. E.g., a NFR may state to minimize the *cost* mandatorily with a maximum of 100 Euros per month.

The three selection phases are:

- *Normalization*. This first phase has the purpose of integrating the potentially heterogeneous QoS attributes' values over which relies decision-making in the WS selection problem. Both inputs WSlist and lreqs must be normalized to compensate the different measurement units of the different QoS values, by projecting them in the interval [0, 1]. Let's denote the normalized structures by WSlistN and lreqsN.
- *Ranking*. Starting from the data normalized in the previous stage, a ranking algorithm can be applied with the goal of computing some similarity measure among the NFRs in lreqsN and the QoS of each service (rows of WSlistN). This algorithm may be any of the commonly used Vector Space Models employed in the context of evaluation of similarity between two objects described by vectors [14]. For example, the Euclidian distance algorithm (see below) looks for the shortest distances between the QoS of each candidate WS and the user NFRs. As a result, we obtain the values of the algorithm and the WS ranked according to them.

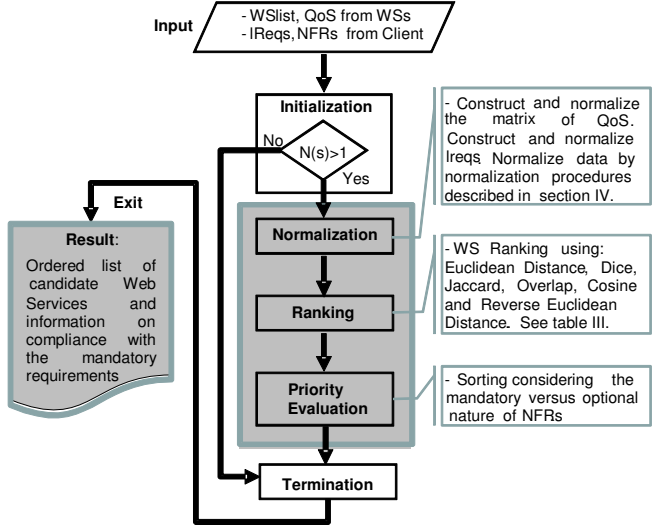

**Figure 1.** Flow diagram for the web services selection process.

- *Priority evaluation.* Last, information about mandatory requirements (stated in lreqtsN) is used. As main ordering criterion, WS are ordered by number of mandatory requirements fulfilled and then, for each category, the results of the ranking algorithm are applied.

## IV. NORMALIZATION AND RANKING ALGORITHMS

As mentioned above, the WeSSQoS architecture supports the coexistence of several normalization and ranking algorithms that are offered as services (see section V).

The normalization service currently offers four normalization algorithms that when applied over a vector $A = (a_1, a_2, a_3, \ldots, a_n)$ that represents quality attributes' values (either from NFRs or QoS), behave as stated in Table II, being $A_i$ the normalization of the value $a_i$ according to the rest of the information encoded in the vector $A$. The comparative analysis of these algorithms (e.g., in terms of preservation of proportionality and clustering of values) is out of the scope of the paper, see e.g. [15] for details.

TABLE II. NORMALIZATION ALGORITHMS OFFERED BY WeSSQoS.

| (1) $A_i = \dfrac{a_i}{Max\ a_i}$ | (3) $A_i = \dfrac{a_i - Min\ a_i}{Max\ a_i - Min\ a_i}$ |
|---|---|
| (2) $A_i = \dfrac{a_i}{\sum_{i=1}^{n} a_i}$ | (4) $A_i = \dfrac{a_i}{\sqrt{\sum_{i=1}^{n} a_i^2}}$ |

On the other hand, the ranking service currently offers six ranking algorithms that when applied over two vectors $A = (a_1, a_2, a_3, \ldots, a_n)$ and $B = (a_1, a_2, \ldots, a_n)$ that represent NFR and QoS such that each pair $(a_i, b_i)$ refers to the same quality attribute $q_i$, behave as stated in Table III. Again details of these six algorithms are out of the scope of the paper, see [9][16] for details.

We present next an example that shows the functioning of the WeSSQoS selection process. Table IV shows at the first row the input to the process and thus to the first stage,

*Normalization.* The second row shows the result of normalization when applying the normalization algorithm (1) from Table II. The third column of Table V shows the result of the *Ranking* stage that uses the second row of Table IV as input. The ranked results using the Euclidean distance algorithm are shown in the fourth column of Table V. It may be observed that WS1 has the minimum value, thus it looks like a promising candidate for selection before evaluating the compliance degree of the mandatory requirements.

TABLE III. RANKING ALGORITHMS OFFERED BY WeSSQoS.

| Similarity cosine | $Sim_{cos\ eno}(A,B) = \dfrac{\sum_{i=1}^{n}(a_i * b_i)}{\sqrt{\sum_{i=1}^{n} a_i^2 * \sum_{i=1}^{n} b_i^2}}$ |
|---|---|
| Jaccard coefficient | $Sim_{jaccard}(A,B) = \dfrac{\sum_{i=1}^{n}(a_i * b_i)}{\sum_{i=1}^{n} a_i + \sum_{i=1}^{n} b_i - \sum_{i=n}^{n}(a_i * b_i)}$ |
| Overlap coefficient | $Sim_{overlap}(A,B) = \dfrac{\sum_{i=1}^{n}(a_i * b_i)}{Min(\sum_{i=1}^{n} a_i, \sum_{i=1}^{n} b_i)}$ |
| Euclidean distance | $Dist_{euclidean}(A,B) = \sqrt{\sum_{i=1}^{n}(a_i - b_i)^2}$ |
| Dice coefficient | $Sim_{dice}(A,B) = \dfrac{2 * \sum_{i=1}^{n}(a_i * b_i)}{\sum_{i=1}^{n} a_i + \sum_{i=1}^{n} b_i}$ |
| Inverse Euclidean distance | $InvDist_{euclidean}(A,B) = \dfrac{1}{1 + \sqrt{\sum_{i=1}^{n}(a_i - b_i)^2}}$ |

As for the Mandatory Evaluation stage, let's suppose that all requirements are mandatory. In the fifth column of Table V, it is shown that WS1 and WS2 comply with 5 out of the 8 NFRs, while WS3 and WS4 comply with 3. If results are combined (with the mandatory requirement compliance set as priority and then the QoS value, the prioritized list of WS changes (see sixth column of Table V) but still WS1 is the best ranked.

TABLE IV. BEFORE AND AFTER THE GENERATION PHASE.

| NFRs from Stakeholder, lreqs: | QoS from candidate WS, WSlist | | | | | | | |
|---|---|---|---|---|---|---|---|---|
| [30, 35, 31, 15, 20, 0.5, .03, 150] | WS1 | 20 | 30 | 25 | 15 | 10 | 0.4 | 0.3 | 50 |
| | WS2 | 5 | 10 | 20 | 20 | 15 | 0.5 | 0.2 | 80 |
| | WS3 | 33 | 11 | 6 | 8 | 10 | 0.8 | 0.4 | 125 |
| | WS4 | 25 | 35 | 45 | 45 | 15 | .5 | .5 | 302 |
| Normalized NFRs, lreqsN: | Normalized QoS from candidate WS, WSlistN | | | | | | | |
| [.91, 1, .69, .33, 1, .62, .60, .50] | WS1 | 0.67 | 0.86 | 0.56 | 0.33 | 0.50 | 0.50 | 0.6 | 0.16 |
| | WS2 | 0.12 | 0.28 | 0.44 | 0.44 | 0.75 | 0.62 | 0.4 | 0.26 |
| | WS3 | 1 | 0.31 | 0.13 | 0.178 | 0.50 | 1.00 | 0.8 | 0.41 |
| | WS4 | 0.76 | 1 | 1 | 1 | 0.75 | 0.62 | 1 | 1 |

TABLE V. RESULTS FROM RANKING AND MANDATORY EVALUATION.

| ID | Name WS | Euclidian distance | Ordering by QoS | Mandatory | Cross-priority QoS/Mandatory |
|---|---|---|---|---|---|
| WS1 | AirportWeatherCheck | 0.71083 | 1 | 5/8 | 1 |
| WS2 | BerreWeather | 1.14562 | 4 | 5/8 | 2 |
| WS3 | fastweather2 | 1.11749 | 3 | 3/8 | 4 |
| WS4 | Weather | 1.01981 | 2 | 3/8 | 3 |

## V. GENERAL DESCRIPTION OF WeSSQoS ARCHITECTURE

The *WeSSQoS* system is structured as an SOA in order to facilitate its integration into other systems. The elements of this architecture are (see Fig. 2):

- *QoSSelector*. This service integrates three services: *QoSRepositoryProxy*, *QoSNormalizeData* and *QoSSelectionModel* (see below), providing a unified view and a single entry point to the whole system.
- *QoSRepositoryProxy*. This is a service that obtains the QoS of WS that belong to a given domain. Two sources of QoS information are defined:
  - *Monitor*. Obtains the QoS at execution time by means of monitoring techniques. As described in the introduction, a monitor works on a predefined catalog of dynamic attributes. No information on static quality attributes will be available, e.g., information on cost of service.
  - *Data Bank*. Obtains the QoS from the provider. QoS is described in an extended WSDL. In the case of dynamic QoS attributes, such as *mean response time*, the QoS value is the one that the provider promises to deliver.
- *QoSNormalizeData*. It normalizes the stakeholder requirements and the QoS data obtained from the Web services by applying normalization procedures, as described in Section IV. Its SOA is flexible enough as to extend the portfolio of normalization algorithms. In its current version, WeSSQoS supports four normalization algorithms. The user provides as input the choice of algorithm.

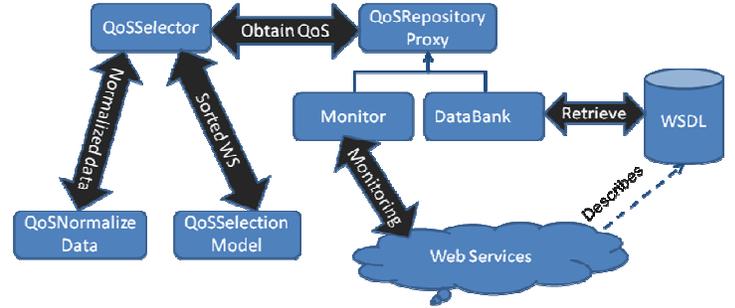

**Figure 2.** WeSSQoS General Architecture.

- *QoSSelectionModel*. It orders the candidate WS by applying ranking algorithms, such as those described in Section IV. Same as the *QoSNormalizeData* service, this service internal architecture is flexible enough as to extend the portfolio of ranking algorithms. Currently, this architecture supports six ranking algorithms. This service also receives as input the user choice of ranking algorithm.

Figure 2 shows the services that support the functionality of WeSSQoS. The basic use case is the ranking of WS represented by the *QoSSelector* service, whose sequence diagram is shown in Figure 3. The operation *rank4QoSRepository* delivers a list of WS for which there is information in the repositories, ranked according to their satisfaction of NFR and mandatory nature, according to the process described in Section III. Figure 4 shows the interfaces of the services, whilst attributes and classes involved are represented in Figure 5.

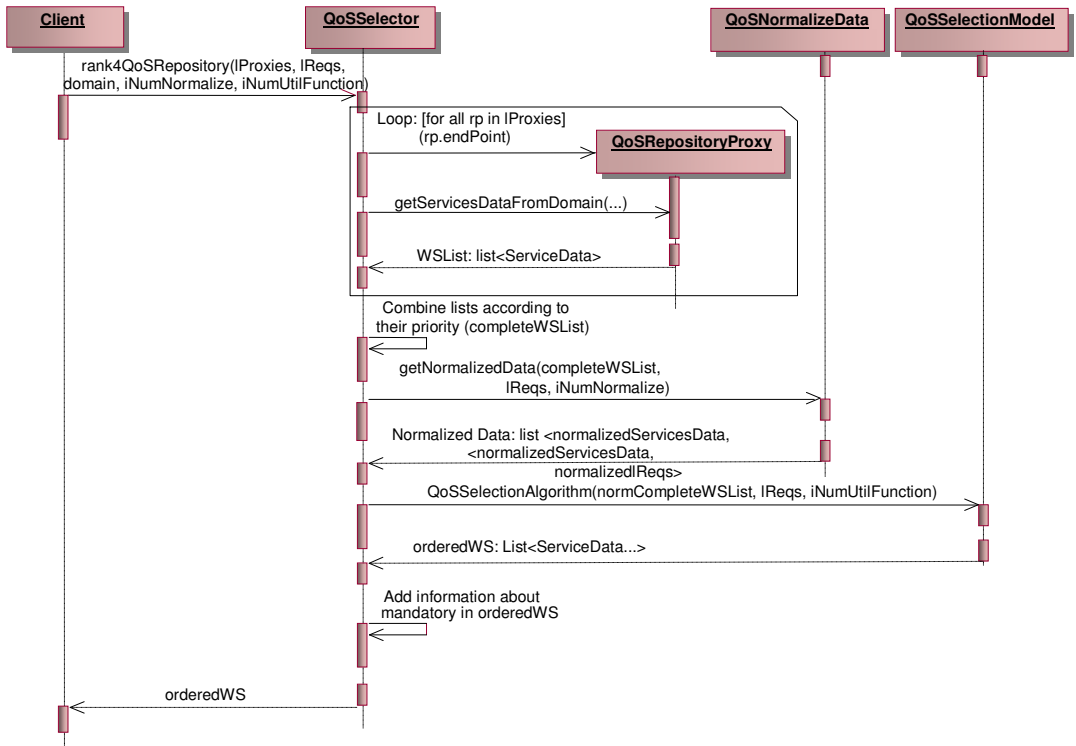

**Figure 3.** Sequence diagram from the ranking WS use case.

| QoSSelector | QoSRepositoryProxy | QoSNormalizeData | QoSSelectionModel |
|---|---|---|---|
| **Operation**: rank4QoSRepository **Input parameters:** lProxies: list <RepositoryProxy> lReqs: list <Stakeholder Requirements> domain: string iNumNormalize: int iNumUtilFunction: int | **Operation:** getServicesDataFromDomain **Input parameters:** domain: string | **Operation:** getNormalizedData **Input parameters:** completeWSList: list <ServiceData> lReqs: list <Stakeholder Requirements> iNumNormalize: int | **Operation:** QoSSelectionAlgorithm **Input parameters:** CompleteWSList: list<ServiceData> lReqs: list <Stakeholder Requirements> iNumUtilFunction: int |
| **Result**: orderedWS: list <ServiceData PriorityResult> | **Result**: WSList: list <ServiceData> | **Result**: NormalizedData: list <normalizedServiceData, normalizedlReqs> | **Result**: orderedWS: list <ServiceDataPriorityResult> |

**Figure 4.** Interfaces of the WeSSQoS services.

- The list of repositories *lproxies* is a list from which the *QoSRepositoryProxy* service obtains the QoS Data. Each repository is represented by its name, the endpoint that allows a service obtaining information (the endpoints correspond to the addresses where the repositories are, either Databanks or Monitors) and their description. Endpoints act as identifiers.
- Each requirement from the NFR list *lReqs* is represented by the name of the quality attribute, the value that is required from the attribute, and the two Boolean values already introduced in Section III (maximize vs. minimize, mandatory vs. non-mandatory).
- The software *domain* is represented by a name and it allows classifying WS into classes.
- The identifiers *iNumNormalize* and *iNumUtilFunction* represent the normalization and ranking algorithms, respectively.

In Figure 3, it is shown that the *rank4QoSRepository* operation calls the *getServicesDataFromDomain* operation for each of the *QoSRepositoryProxy* (*Databank* or *Monitor*) specified in *lproxies*. The endpoint is used at the moment of creating the *QoSRepositoryProxy*. While obtaining the list of the services, the list WSlist with the QoS information from the NFR needed is prepared. In the case of having information about a NFR in more than one repository, a simple priority policy is applied: the information that appears first in one repository from the repositories list is taken into account (in other words, the order in the list of repositories determines the priority of the attributes that are in more than one repository). Once the list WSlist of services and their QoS is obtained, the three steps presented in Fig. 1 are applied. First, the *getNormalizedData* operation from the *QoSNormalizedData* service is executed taken as input the *WSlist*, the user NFR represented by *lreqs* and the selected type of normalization process represented by *iNumNormalize* in order to normalize the data from the services and from the user. Afterwards, the operation *QoSSelectionAlgorithm* of the *QoSSelectionModel* service is executed in order to rank the WS with the algorithm identified by *iNumUtilFunction*. Finally, a ranked list of WS is produced considering priority of attributes.

Figure 5 sums up the interfaces that appear in the sequence diagram.

## VI. PROTOTYPE DESCRIPTION

The WeSSQoS system described so far is implemented and available in http://appserv.lsi.upc.es/wessqos/. The system has been developed using Java J2EE and Apache Axis2 as web service technology, and Apache Tomcat as execution platform. A tested of WS belonging to different domains and placed in different repositories has been developed for testing purposes, using the Glassfish web service technology in order to assess technological independence of the platform.

We have created a client web interface divided in several areas described next (see Figure 6). The first area, *Repositories*, is used to define the domain and the repositories over which the search will be done. The system allows using both internal repositories (i.e., local to WeSSQoS) and external ones. The domain name is required to identify the required services in selected repositories. On the other hand, the repositories are identified using their endpoint. The endpoints are shown in the table at the end of this area, being possible to add and remove them from the repository list.

The second area renders the available portfolios of normalization and ranking algorithms so that the analyst may select one.

The third area, Stakeholder Requirements, is the section where the analyst introduces the NFR to be fulfilled. These NFR are settled over quality attributes that can be the attributes provided by WeSSQoS system itself (coming from [4]) or by the analyst. Clearly, the analyst has the responsibility of choosing quality attributes which description or monitoring is covered by the repositories chosen on Repositories section. For each selected attributes, it must be introduced the required value, if this attribute must be minimized or maximized and if it is mandatory to select services (as is described in Section V).

Finally, the *Results* area shows the resulting ranking. The ranking is a sorted WS list according to the ranking algorithm chosen by a user. Figure 7 shows the results of the ranking phase using the Euclidean Distance. Then a user may also rank the WS list according to the mandatory requirements. This is shown in Figure 8.

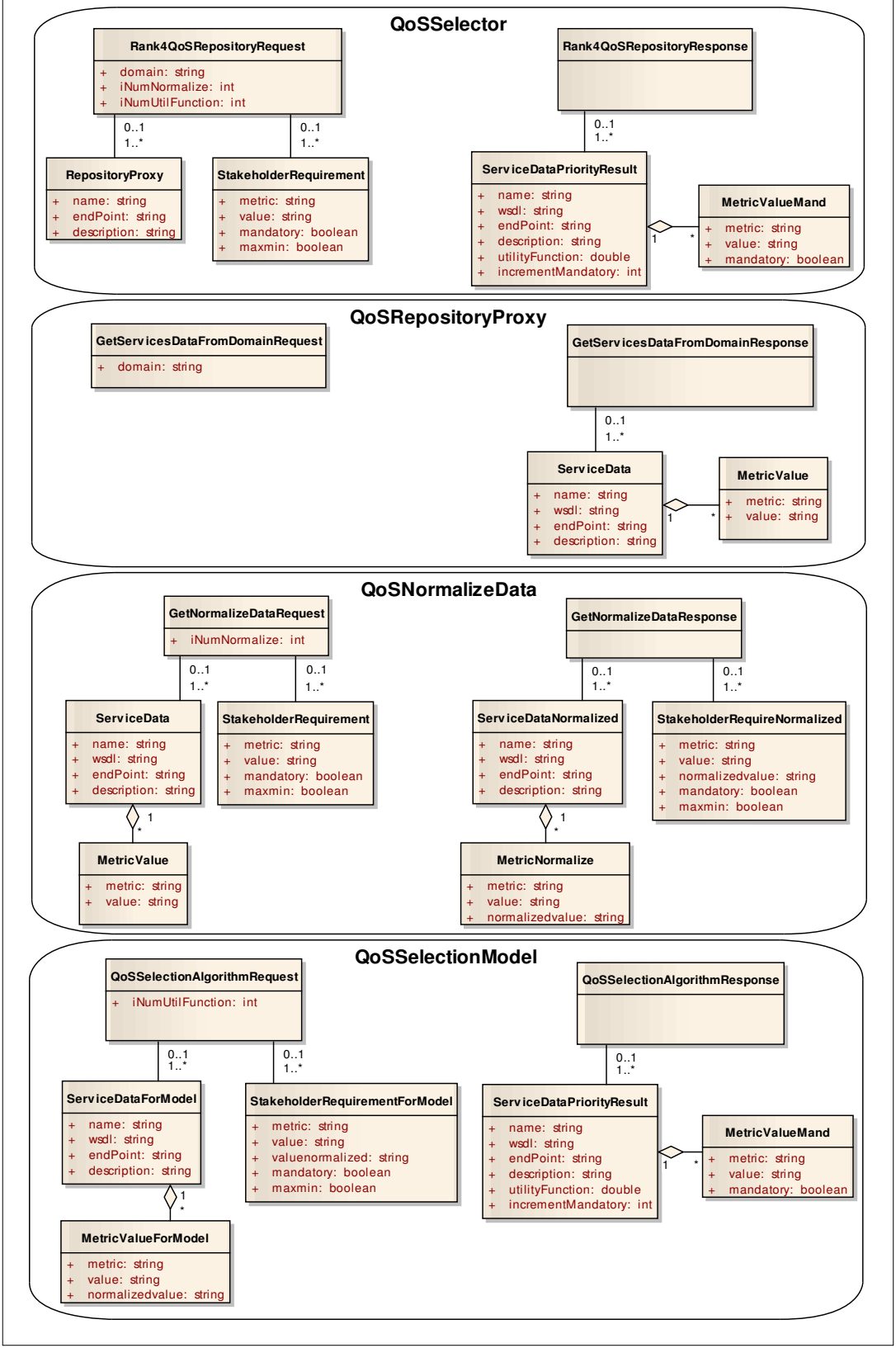

**Figure 5.** Class Diagram of the Internal Architecture of the WeSSQoS services.

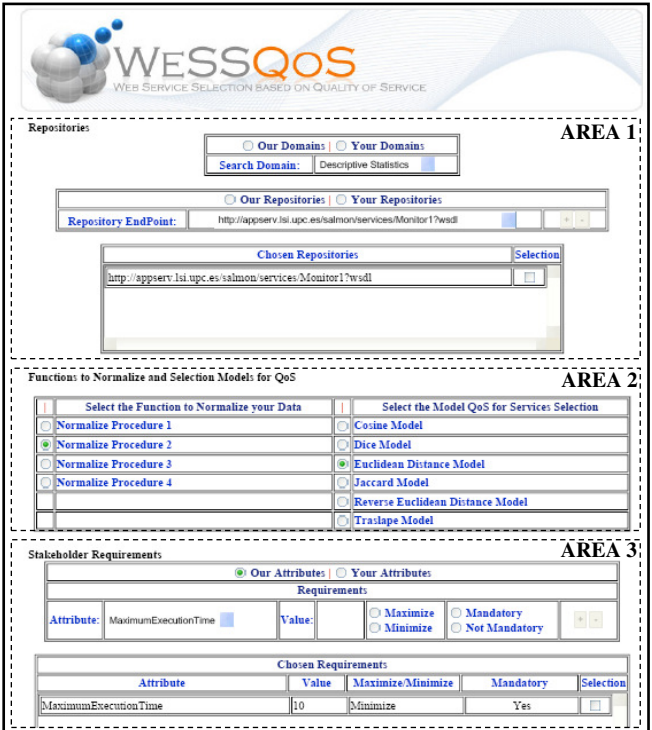

**Figure 6.** WeSSQoS' snapshot (see http://appserv.lsi.upc.es/wessqos/)

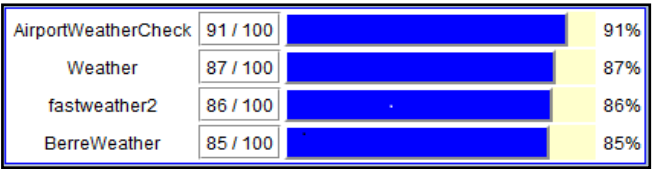

**Figure 7.** Ranking Results using Euclidean Distance.

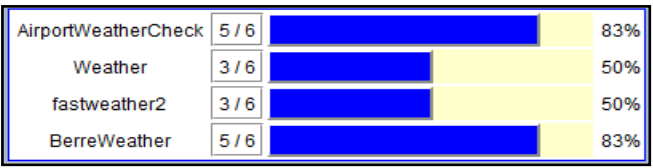

**Figure 8.** Ranking Results considering Mandatory Requirements.

## VII. VALIDATION

In order to test our tool, we have design a scenario to execute some test cases. This scenario is also available in the prototype (see Section VI). It has been designed to test the following features of our system:

- Quality attributes management. The analyst can decide the quality attributes which she is interested on. These attributes can be or not defined in the information about the WS we are selecting. The basic case is when the customer asks for a subset of the attributes defined on the repositories. The customer can also ask for attributes that are not specified on repositories, these attributes will be treated as undefined by the ranking algorithm.
- Repositories independence. Our system does not have restriction in the number or repositories used for the search. Each repository can be static or dynamic. When there is more than one repository, the following assumptions have been taken:
  o The WS of each repository can be different. In this case we consider as WS candidates the union of all services inside all repositories.
  o More than one repository may contain information of a given WS, but the quality attributes are disjoint. In this case, the algorithm will simply combine the required attributes getting them from the adequate repositories.
  o More than one repository may contain information of a given WS, and some quality attribute may appear in more than one repository. In this situation, the value is taken from the repository with more priority (i.e., the one declared first).

Figure 9 shows the architecture of our implementation and the necessary data for running the tests previously described. We have both types of *QoSRepositoryProxy* (static and dynamic). The *Monitor* instances use Axis, whilst the *DataBank* (which contains information about two WS domains) uses Glassfish. Next to the *QoSRepositoryProxy* has been included the name of some of its WS. These services have been selected in order to highlight services located in more than one repository, and some of them have attributes also in more than one repository. In the *Databank*, there is information about all attributes with the exception of *CurrentResponseTime* (CRT) and *CurrentAvailability* (CA). In the *Monitor* services, the information about what attributes have information are included too. In addition to the CRT and CA, there is also information about the *AverageResponseTime* (ART) in some services. If the priority of repositories (i.e., their order of appearance) is Monitor1, Monitor2, DataBank1, given the service *AirportWeatherCheck* (which is located in all the repositories) ART, CRT and CA will be taken from the Monitor1 and the other attributes from the DataBank1. However, if the order was Monitor2, Monitor1 y DataBank1, the CRT would be taken from the Monitor2, ART and CA from the Monitor1 and the rest from the DataBank1.

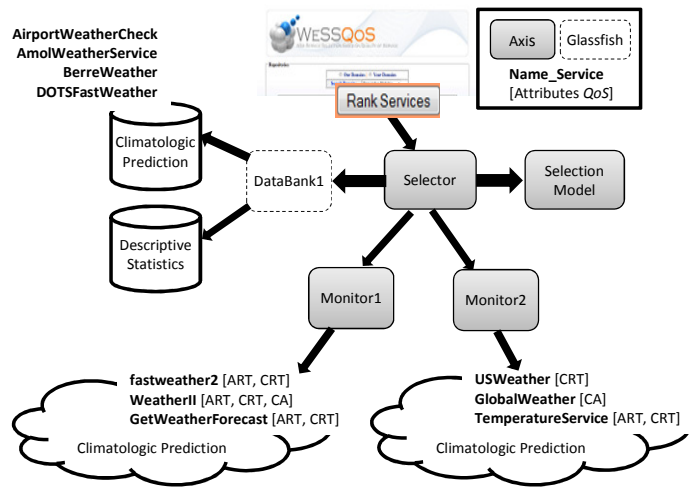

**Figure 9.** Scenario for WeSSQoS tests.

## VIII. CONCLUSIONS AND FUTURE WORK

In this paper we have presented WeSSQoS, a framework for ranking available WS through the evaluation of their QoS with respect to the stated NFRs. In terms of the criteria introduced in Section II, we can conclude that our proposal has the following advantages:

- *Architectural style*. WeSSQoS is developed as a Service Oriented System itself. Following SOA principles, users can add new services (new ranking algorithms services, new repositories,…) if they are just compliant with the expected service definitions (which are described in section VI).
- *Quality attributes*. WeSSQoS is independent of the Quality Model or ontology used to define quality attributes. The interface of the system allows the user to select from a well-known predefined set of attributes based on [4], and also add any kind of quality attributes from any Quality Model. As many frameworks, WeSSQoS is able to work with either static or dynamic quality attributes, although it's important to mention that this distinction is implicit from the way the data has been retrieved.
- *QoS Data*. WeSSQoS is able to retrieve quality attributes from either quality descriptions in the Service definition or by monitoring systems. The usage of a common interface (Proxy) to retrieve the data in a uniform way from these sources, provides extensibility to add new kind of repositories, independently of the approach used to obtain the data.
- *Multialgorithm*: WeSSQoS is able to work with any kind of ranking algorithm that is implemented using the defined interface. Eventually, we could use arbitrarily complex algorithm, e.g. aggregators of results through choreography of other WS that define different algorithms.
- *Multirepository*: WeSSQoS allows the user to include several repositories of WS with independence of the technology used. Furthermore, it provides a mechanism to combine the QoS data when the same service is present in more than one repository. Currently, the user is responsible to select those repositories that are compatible with each other. (e.g.: repositories should use a common terminology to refer to the same quality attribute)
- *Prototype available*: A prototype of WeSSQoS is available at http://appserv.lsi.upc.es/wessqos/. The current version has been tested and validated as explained in Section VII.

In Section V we dealt with the issue concerning WS repositories' priority policy, the main idea of this is to integrate in WeSSQoS general repository the WS from all chosen repositories in a prioritised way. Therefore, WS integration is used in repositories combination and it is not part of WS composition, we do not treat composition topics in the paper.

As future work, we have identified several research lines and improvements that could be performed in order to increase the current framework's capabilities:

- Perform tests in large web service ecosystems to ensure the correctness and suitability of the framework to rank web services in real situations.
- Increase the number of dynamic quality attributes retrieved by the monitoring system.
- Design different sophisticated mechanisms to combine data from several repositories (and unify these strategies under a common interface, in order to build it as a service)

## ACKNOWLEDGEMENTS

This work has been partially funded by the Spanish project TIN2010-19130-C02-01. Oscar Cabrera has completed a stay at the UPC using a CONACYT grant.